\documentclass[%
 aip,
 amsmath,amssymb,
 reprint,%
 floatfix,
]{revtex4-1}

\usepackage{graphicx}
\usepackage{dcolumn}
\usepackage{bm}
\usepackage[export]{adjustbox}
\usepackage[utf8]{inputenc}
\usepackage[T1]{fontenc}
\usepackage{mathptmx}
\usepackage{etoolbox}
\usepackage{placeins}
\usepackage{multirow}
\usepackage[caption=false]{subfig} 

\makeatletter
\def\@email#1#2{%
 \endgroup
 \patchcmd{\titleblock@produce}
  {\frontmatter@RRAPformat}
  {\frontmatter@RRAPformat{\produce@RRAP{*#1\href{mailto:#2}{#2}}}\frontmatter@RRAPformat}
  {}{}
}%
\makeatother
\begin{document}

\preprint{AIP/123-QED}

\title{Efficient optical coupling to gallium arsenide nano-waveguides and resonators with etched conical fibers}

\author{S. Pautrel}
\affiliation{Matériaux et Phénomènes Quantiques (MPQ), Université Paris Cité, CNRS, UMR 7162, 75013 Paris, France}%
\author{F. Malabat}
\affiliation{Matériaux et Phénomènes Quantiques (MPQ), Université Paris Cité, CNRS, UMR 7162, 75013 Paris, France}%
\author{L. Waquier}
\affiliation{Matériaux et Phénomènes Quantiques (MPQ), Université Paris Cité, CNRS, UMR 7162, 75013 Paris, France}%
\author{M. Colombano}
\affiliation{Matériaux et Phénomènes Quantiques (MPQ), Université Paris Cité, CNRS, UMR 7162, 75013 Paris, France}%
\author{M. Morassi}
\affiliation{Centre de Nanosciences et de Nanotechnologies, Université Paris-Saclay, CNRS, UMR 9001, 91120 Palaiseau, France}
\author{A. Lema\^itre}
\affiliation{Centre de Nanosciences et de Nanotechnologies, Université Paris-Saclay, CNRS, UMR 9001, 91120 Palaiseau, France}
\author{I. Favero$^*$}
\affiliation{Matériaux et Phénomènes Quantiques (MPQ), Université Paris Cité, CNRS, UMR 7162, 75013 Paris, France}%

 \email{ivan.favero@u-paris.fr}

\date{\today}

\begin{abstract}
We explore new methods for coupling light to on-chip gallium arsenide nanophotonic structures using etched conical optical fibers. With a single-sided conical fiber taper, we demonstrate efficient coupling to an on-chip photonic bus waveguide in a liquid environment. We then show that it is possible to replace such on-chip bus waveguide by two joined conical fibers in order to directly couple light into a target whispering gallery disk resonator. This latter approach proves compliant with demanding environments, such as a vibrating pulse tube cryostat operating at low temperature, and it is demonstrated both in the telecom band and in the near infrared close to 900 nm of wavelength. The versatility, stability, and high coupling efficiency of this method are promising for quantum optics and sensing experiments in constrained environments, where obtaining high signal-to-noise ratio remains a challenge.\end{abstract}

\maketitle

\section{Introduction}

Tapered optical fibers are versatile tools used to probe photonic devices \cite{Knight1997}, but also to study cavity electrodynamics with quantum dots \cite{Srinivasan2007}, with optomechanical resonators \cite{Carmon2005, Ding2010HighFrequency, Ding2010LoopedFiber}, with a single atom \cite{Nieddu2016} or with an array of atoms \cite{Thompson2013}. However, implementing tapered fiber optical coupling schemes in a cryostat for low temperature experiments, sometimes required in these fields, remains technically difficult \cite{riviere_evanescent_2013,srinivasan_optical_2007, MacDonaldDavis2010, Wasserman2022}. 

In several groups, commercial lensed fibers have been used in a cryostat to couple light to on-chip bus waveguides, which themselves couple to the emitter or resonator of interest \cite{Gil-Santos-LightMediated2017, ren_two-dimensional_2020,jiang_optically_2023}.
On-chip coupling waveguide architectures provide additionally an inherent signal stability, since the relative positions of the resonator and waveguide are fixed. In consequence, the interest of on-chip waveguides does not restrict to cryogenic experiments, and the approach has also spread to nanophotonics applications in liquids\cite{Gil-Santos2015}, such as biosensing or rheology. 
However, despite relatively high single-port coupling efficiencies $\eta$ of the order of 50-70$\%$ between a lensed fiber and a nanowaveguide \cite{meenehan_pulsed_2015,patel_room-temperature_2021, Hease2016PhdThesis}, it is often necessary to sweep several parameters on the same chip in order to obtain a satisfying waveguide-to-resonator coupling, limiting the number of useful resonators. On top of this, the integration of a bus waveguide on the chip is not always possible, depending on the specific material or nanofabrication limitations.
Hence for very demanding experiments such as single-photon quantum optomechanics \cite{Pautrel2020}, optomechanics in liquids \cite{Gil-Santos2015}, or enhanced optomechanics with quantum well exciton-polaritons \cite{CarlonZambon2022}, it would be highly desirable not only to enhance $\eta$ but also to gain supplementary degrees of freedom in the coupling method. Recently, a technique using a single-sided conical fiber taper was proposed for efficient coupling to silicon, silicon nitride and diamond optical resonators \cite{Burek2017, Tiecke2015}, which improves some of these aspects. 
\\

We adapt here this method to gallium arsenide nanophotonic structures, varying within the evanescent coupling region the diameter of a conical fiber taper and its coupling distance. 
We then introduce an important variation to the method, where two conical fiber tapers are now joined to form a stable two-port optical coupling device. 
We apply these developments to couple light both to nanoscale suspended waveguides and whispering gallery resonators, and focus on very constraining environments such as a liquid or a pulse-tube dry cryostat operating at low temperature. In theses environments, we demonstrate resonant laser spectroscopy of resonators at a wavelength of 1.5 $\mu$m, as well as off-resonant photoluminescence spectroscopy of InGaAs quantum wells at a wavelength of 850~nm. Finally, we provide full 3D simulations of fiber-to-waveguide and fiber-to-fiber coupling as function of different parameters, such as the cone angle and the contact region of the fibers, and show that it is possible to reach a coupling efficiency above $99\%$ with this approach.

\section{\label{sec:level1}Conical fiber taper coupling to a GaAs nano-waveguide}
\subsection{Integrated nanophotonic device}
\begin{figure*}
\subfloat[]{
        \includegraphics[width=0.48\textwidth]{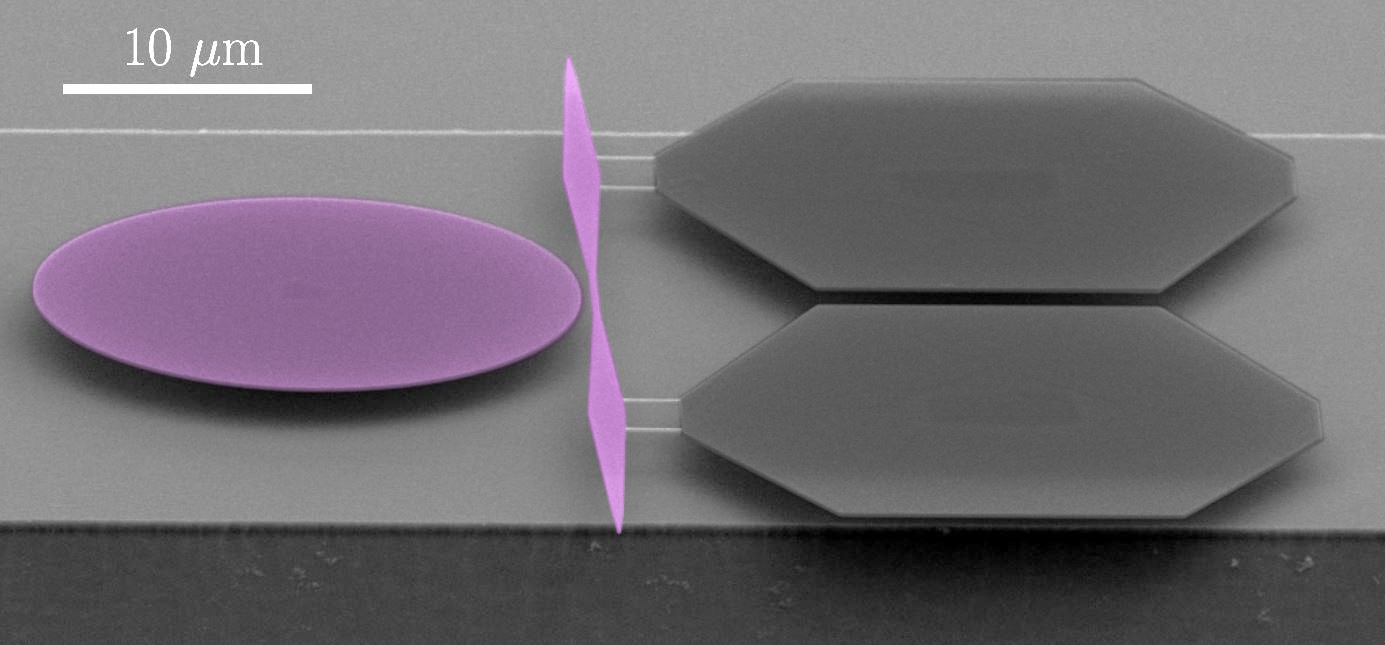}
        \label{fig:fab}
    }
\subfloat[]{
    \includegraphics[width=0.49\textwidth]{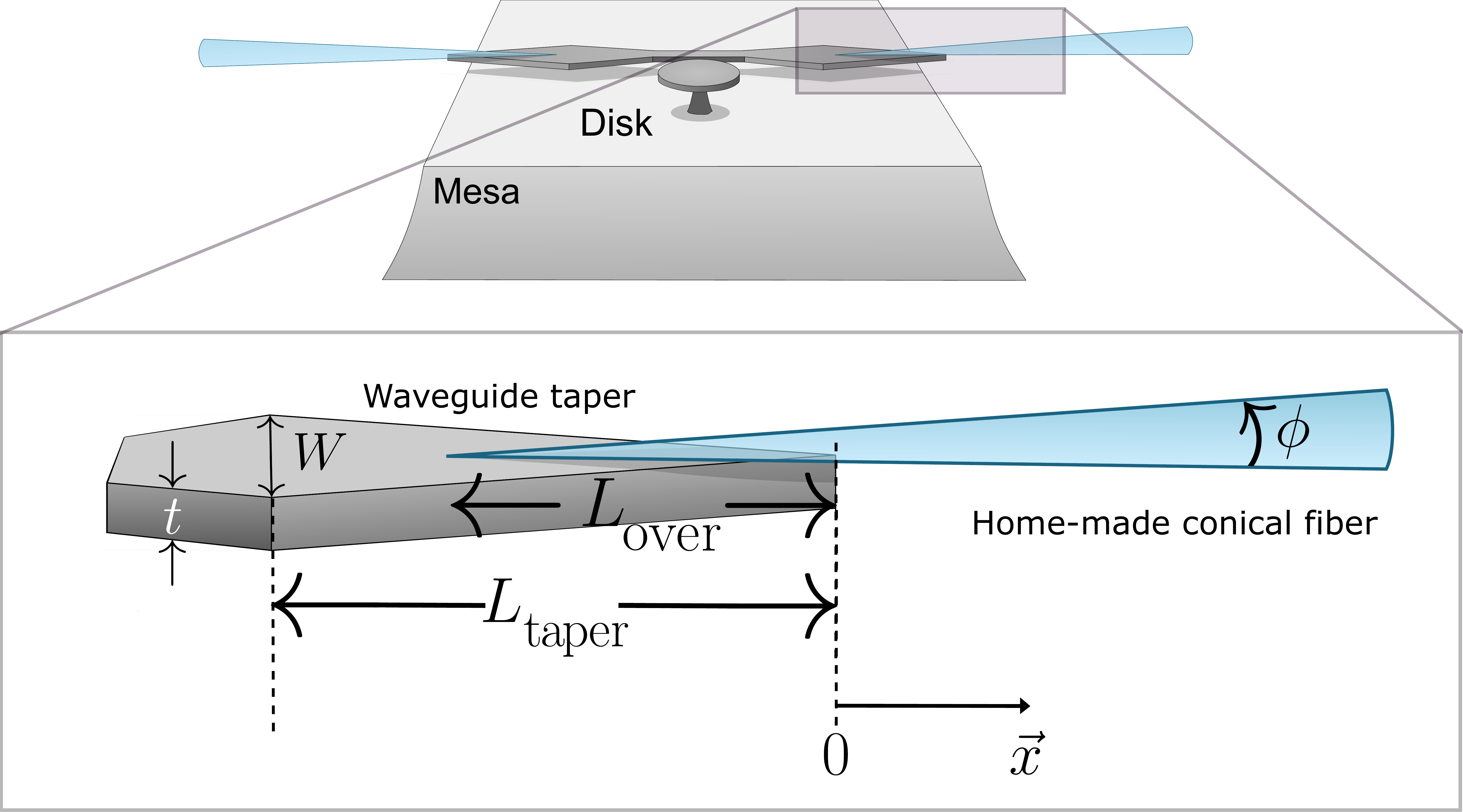}

    \label{fig:coupling_f2wg}
    }\\
\subfloat[]{
    \includegraphics[width=0.985\textwidth]{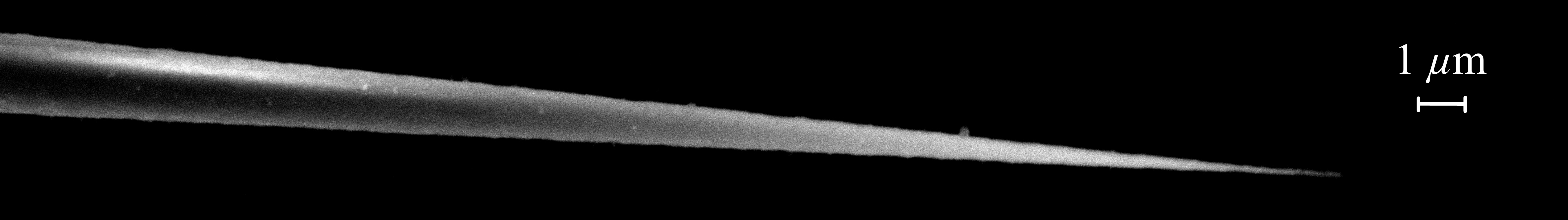}

    \label{fig:fiber_meb}
    }

\caption{
\textbf{(a)} SEM micrograph of a GaAs disk resonator with its fully suspended coupling waveguide (false-colored in purple). The disk radius is 11 $\mu$m, its thickness is 200 nm. The hexagonal anchoring pads supporting the suspended waveguide are shaded. \textbf{(b)} Coupling principle with a home-made conical fiber contacting the top of each tapered end of the coupling waveguide. The inset shows the simulation geometry, where $W$ is the final waveguide width, $t$ its thickness, and $L_{\mathrm{taper}}$ the tapered length. The conical fiber is characterized by its angle $\phi$, and the overlap length is $L_{\mathrm{over}}$. \textbf{(c)} SEM micrograph of a conical fiber fabricated by us. Some surface roughness is visible, and always present with the current fabrication procedure. However, it is experimentally not detrimental to the optical coupling efficiency.
}
\end{figure*}

The system consists of a nanophotonic bus waveguide, whose two endings have been shaped into inverted tapers, in the vicinity of a GaAs disk resonator that exhibits high-quality optical modes ($Q$ up to a few millions \cite{guha_surface-enhanced_2017}). Fig.~\ref{fig:fab} shows a SEM micrograph of a typical device.
The design of this integrated photonic circuit was optimized for light injection and collection in the telecom band (1500-1600 nm), with microlensed fibers in front of the inverted tapers. In order to approach such 60 $\mu$m radius fiber close to the waveguide tapered end, a \textit{mesa} structure was carved chemically by wet etching, elevating the devices over the substrate.

While this design was originally optimized for micro-lensed fibers \cite{Sbarra2021APL}, it can also be used to contact the top surface of inverted tapers with a conical fiber taper (Fig. \ref{fig:coupling_f2wg}). 
A picture of a fabricated single-sided tapered fiber can be found in Fig.~\ref{fig:fiber_meb} and the fabrication procedure is detailed in Appendix.~\ref{app:fabrication}. At 1550 nm, the coupling efficiency between the waveguide and our home-made conical taper fiber is slightly larger than what we reach with a commercial lensed fiber ($\eta_{\mathrm{max}}^{\mathrm{conical}} \approx 65\%, \eta_{\mathrm{max}}^{\mathrm{lensed}} \approx 50\%$). The main advantages of this coupling method are the stability and the time saving in the alignment, particularly in difficult operating environments such as a liquid.

\subsection{Simulation of the coupling efficiency}
\begin{figure}
\subfloat[]{
    \includegraphics[height=0.27\textwidth]{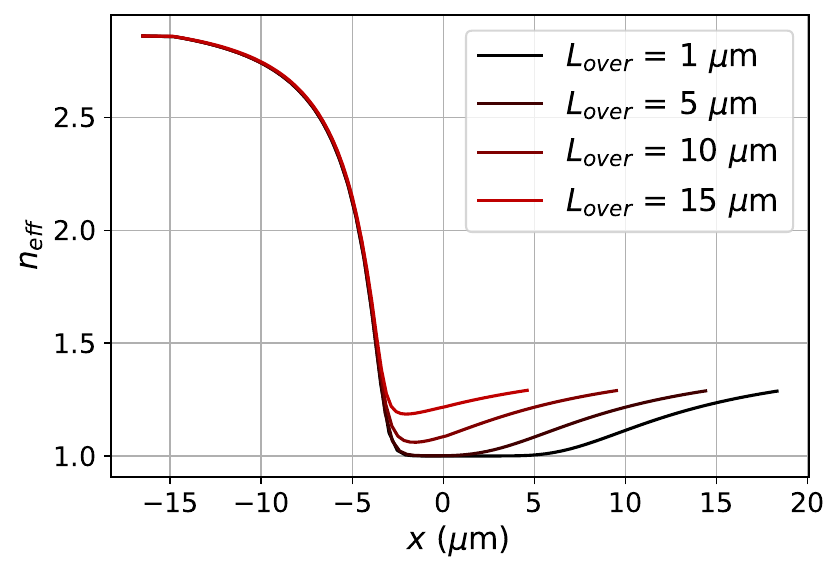}
    \label{fig:simu_GaAsWG_neff}
    }\\
\subfloat[]{
    \includegraphics[height=0.27\textwidth]{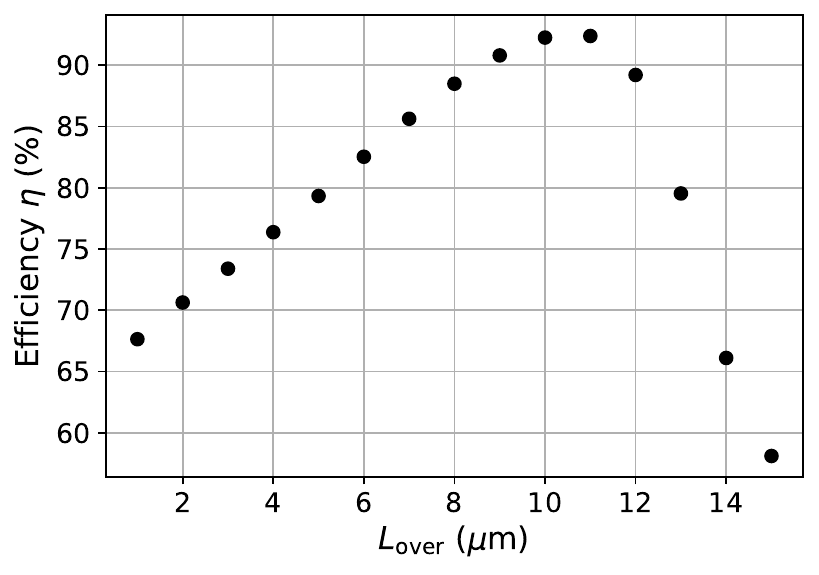}
    \label{fig:simu_GaAsWG_eff}
    }
\caption{
3D EME simulation of the conical fiber taper to GaAs nano-waveguide junction for $\lambda = 1550$ nm. The effective index $n_{\mathrm{eff}}$ of the fundamental supermode along the light propagation axis is shown in  \textbf{(a)} for different values of $L_{\mathrm{over}}$. The coordinate $x=0$ corresponds to the waveguide inverted taper end. In \textbf{(b)}, the simulated single-port power coupling efficiency is shown as a function of $L_{\mathrm{over}}$ The other geometrical parameters are respectively ($W$,$t$,$L_{\mathrm{taper}}$,$\phi$) = (1 $\mu$m, 320 nm, 15 $\mu$m, 4.15$^{\circ}$).
}
\label{fig:simu_GaAsWG}
\end{figure}

We performed a 3D EigenMode Expansion (EME) simulation of the conical fiber taper to waveguide coupling method. The simulation geometry is detailed in the inset of Fig~\ref{fig:coupling_f2wg}. A supended GaAs nano-waveguide of rectangular cross-section with thickness $t$ and nominal width $W$ is linearly tapered over a length $L_{\mathrm{taper}}$ down to a minimal width of 50 nm at its tip.
The waveguide taper geometry is defined independently to respect the adiabaticity criterion \cite{Love1991AdiabaticCriterion} :
\begin{equation}
    L_{\mathrm{taper}} \gg L_{\mathrm{beat}} = \frac{\lambda}{n_{\mathrm{eff},1}-n_{\mathrm{eff},2}}
\end{equation}
where $L_{\mathrm{beat}}$ is the beating length between the first two guided modes (with effective index $n_{\mathrm{eff},1}$ and $n_{\mathrm{eff},2}$ respectively) and $\lambda$ is the free-space wavelength.

A silica cone with a full angle $\phi$ is placed in contact with the top surface of the waveguide tapered region over the overlap length $L_{\mathrm{over}}$.
Simulations results are presented in Fig.~\ref{fig:simu_GaAsWG_neff} and Fig.~\ref{fig:simu_GaAsWG_eff} for specific values of interest $t$, $w$, $L_{\mathrm{taper}}$ and $\phi$ employed in our experiments. Fig.~\ref{fig:simu_GaAsWG_neff} shows the effective index $n_{\mathrm{eff}}$ of the first supermode along the light propagation direction for different values of $L_{\mathrm{over}}$. As expected, a larger overlap length increases the value of $n_{\mathrm{eff}}$, which increases the mode confinement and experimentally reduces scattering-induced losses. Fig.~\ref{fig:simu_GaAsWG_eff} shows the computed coupling efficiency $\eta$ as a function of $L_{\mathrm{over}}$, where a maximum transmission $\eta \approx 92\%$ is found for \mbox{$L_{\mathrm{over}} \approx 11$ $\mu$m}. We do not reach experimentally this level of efficiency yet and scattering loss at the anchoring points of the suspended waveguide are thought to be the main culprit. 

\subsection{Optical coupling in water}

\begin{figure}
    \includegraphics[width=0.48\textwidth]{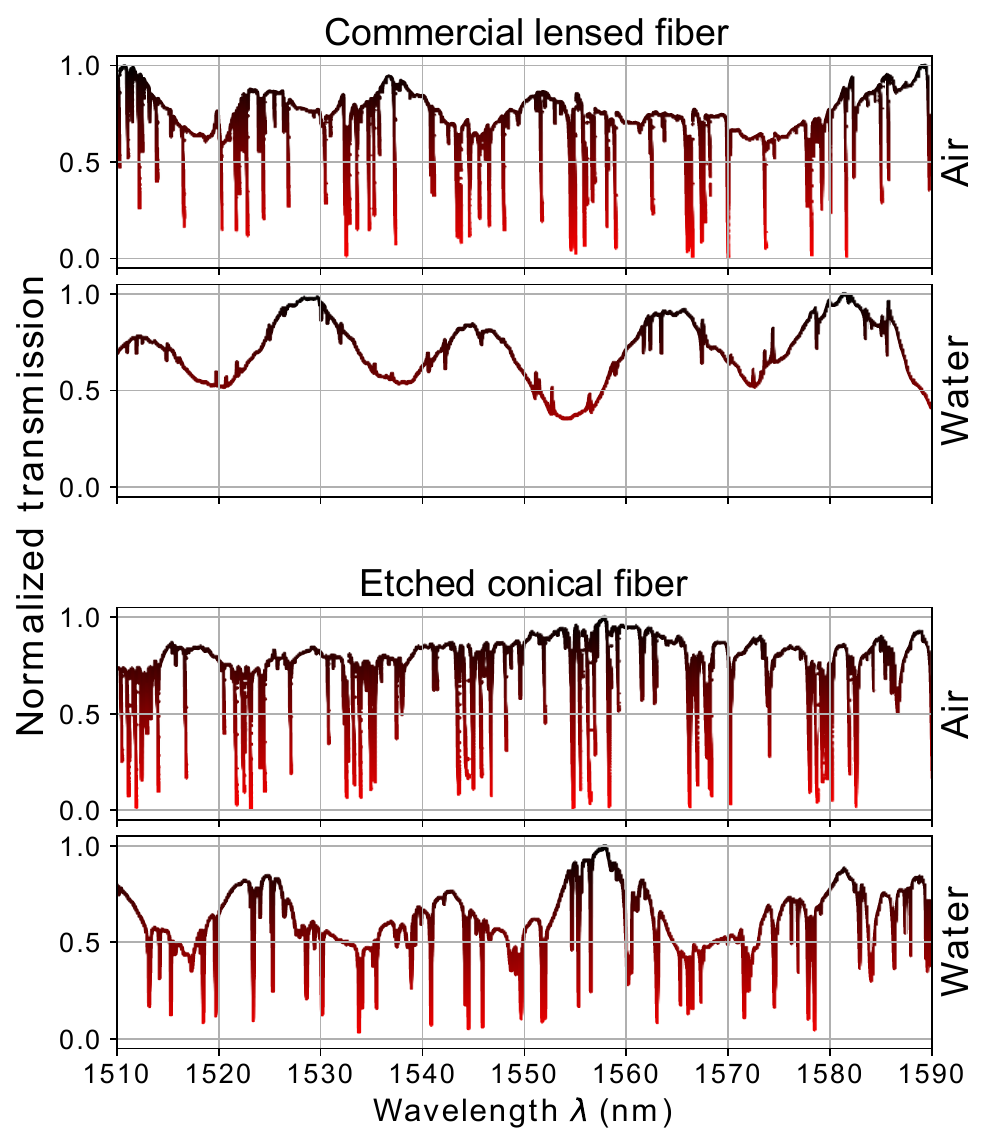}
\caption{Normalized optical spectrum obtained in different conditions for given GaAs disk dimensions (radius 11 $\mu$m, thickness 200 nm) in presence of a coupling waveguide. The spectra in the two top panels have been obtained with commercial lensed fibers, while those of the two bottom panels have been measured with homemade conical fiber tapers brought in contact with the top of the tapered bus waveguide. For each coupling method, spectra obtained in air and in water are shown.}
\label{fig:spectrums_water}
\end{figure}

In a liquid environment (typically water, $n\approx 1.33$), the refractive index of the surrounding medium can be close to that of silica ($n \approx 1.51$) and the light coming from a micro-lensed fiber cannot be focused properly on the tip of the waveguide. As a consequence, we observe an inefficient coupling between commercial micro-lensed fibers and GaAs nano-waveguides. 
Moreover, the refractive index contrast between silica and water renders these lensed fibers almost invisible under an optical microscope.
To overcome these difficulties, we test here the coupling of a GaAs waveguide to conical fibers in water. Fig.~\ref{fig:spectrums_water} provides a comparison of the optical spectrum of a disk resonator in air and in water acquired with commercial lensed fibers or with etched conical fibers in contact with the bus waveguide. With lensed fibers, the spectrum in water features low contrast Fano resonances, characteristic of residual direct fiber-to-fiber coupling and poor injection inside the waveguide. With etched conical fibers in contrast, we retrieve in water optical transmission mode shapes similar to those obtained in air.
Due to van der Waals forces \cite{Marinkovic2021}, the conical fibers stick to the waveguide, which reduces the chances of drift in time. Such property allows long measurement times of hours for biosensing or rheological studies. Additionally, the residual surface roughness of homemade conical fibers, as well as their low curvature radius, make them much more visible, which is convenient for optical alignment.
When moving from air to water, the single-port coupling efficiency with an etched conical fiber drops to $\eta = 25$\%. This is thought to result from a weaker optical mode confinement in water, which makes surface roughness and scattering-induced losses more important. Such single-port coupling efficiency remains slightly higher than what is typically obtained in water with more conventional grating couplers ($\eta \approx 20\%$)\cite{neshasteh_2023}.
\section{Two conical fibers forming a junction: a new bus coupler}
\begin{figure*}

\subfloat[]{
    \includegraphics[width=0.48\textwidth]{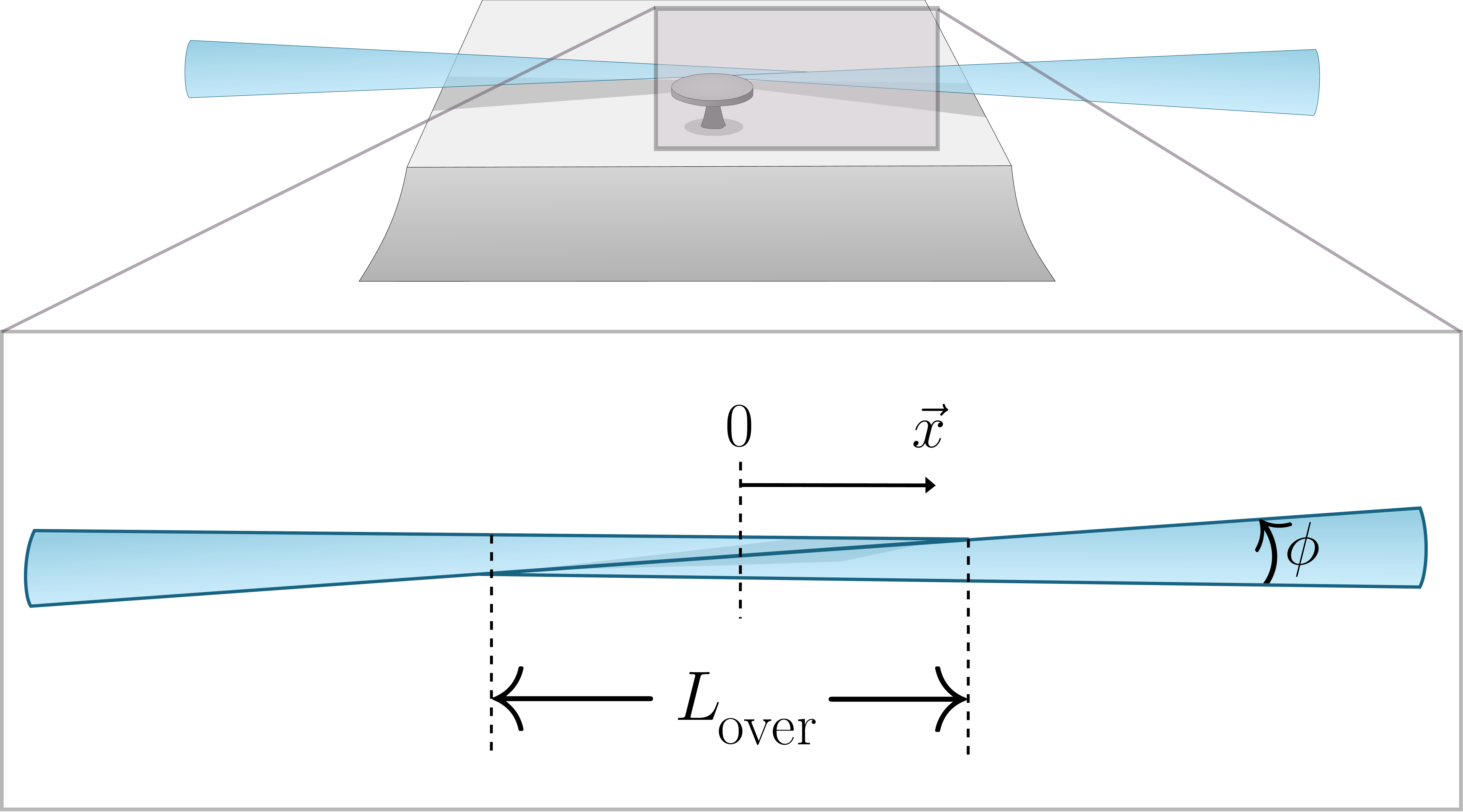}
    \label{fig:coupling_f2f}
    }
\subfloat[]{
    \includegraphics[width=0.43\textwidth]{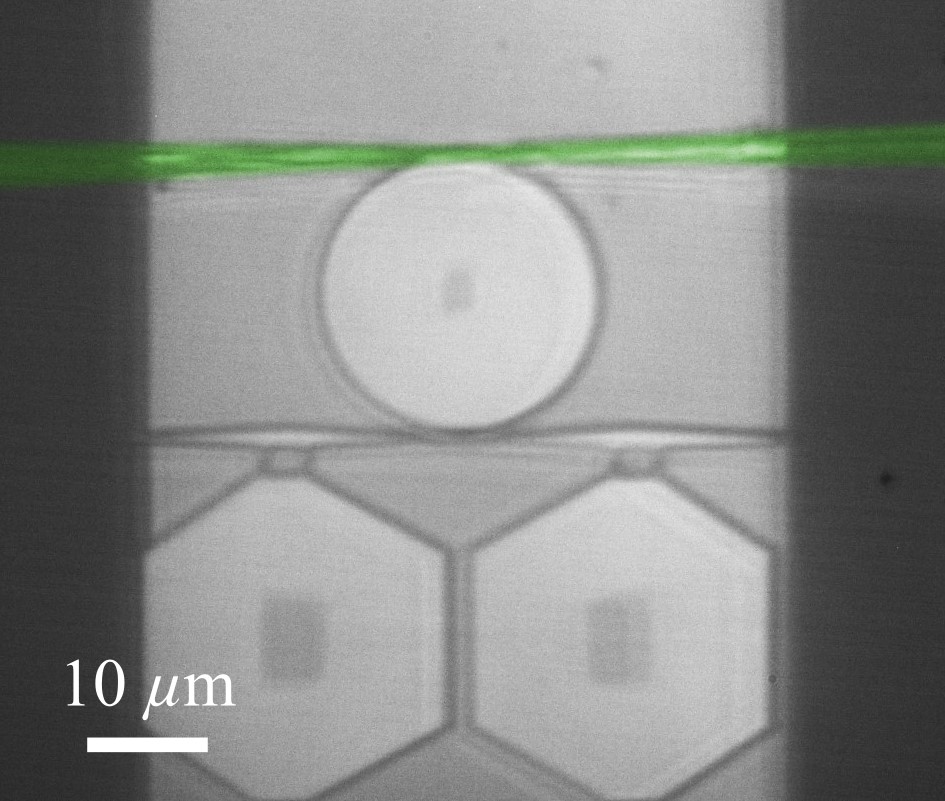}
    \label{fig:image_camera_4K}
    }
\\
\subfloat[]{
    \includegraphics[height=0.25\textwidth]{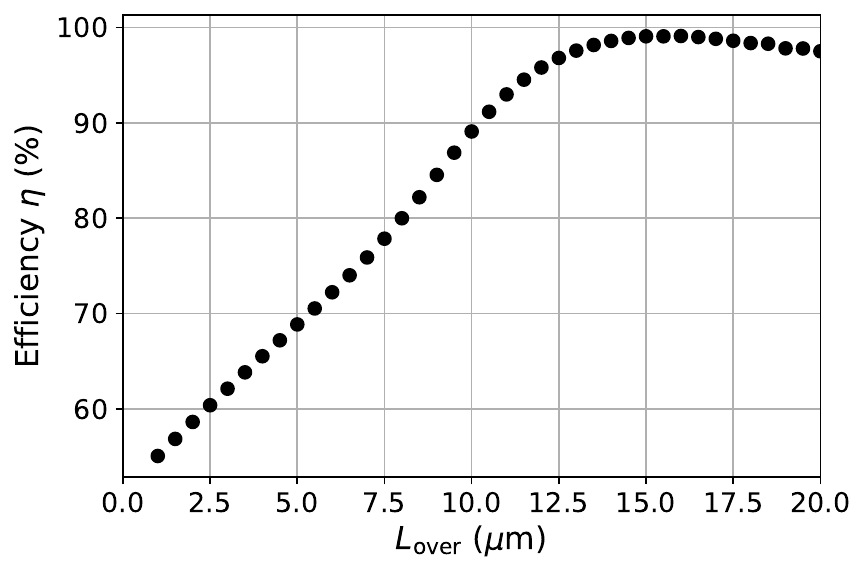}

    \label{fig:simu_f2f_eff}
    }
\subfloat[]{
    \includegraphics[height=0.24\textwidth]{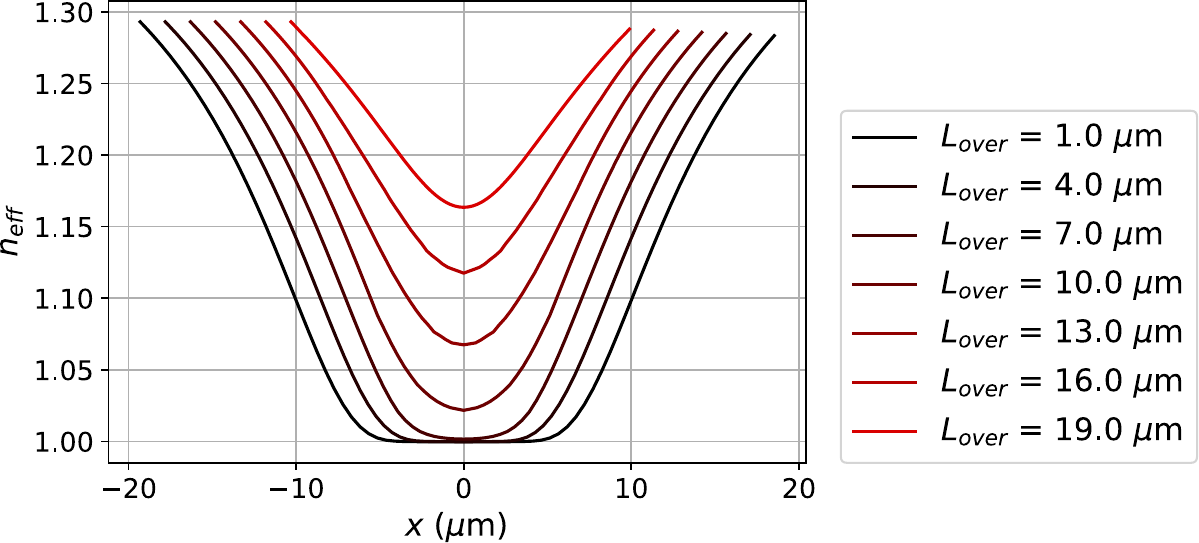}

    \label{fig:simu_f2f_neff}
    }

\caption{
\textbf{(a)} Sketch of the coupling principle with a junction. $x=0$ corresponds to the center of the junction. \textbf{(b)} Experimental situation with a 11 $\mu$m radius, 200 nm thick GaAs disk resonator. The homemade conical fibers forming a junction are false-colored in green (top). A coupling nano-waveguide in GaAs (bottom) is visible but is bypassed in the present experiment with the fiber-fiber junction. \textbf{(c),(d)} 3D EME simulation of a fiber-to-fiber junction as a function of the overlap length $L_{\mathrm{over}}$ for $\lambda = 1550$ nm. \textbf{(c)} shows the simulated power coupling efficiency as a function of $L_{\mathrm{over}}$. The fiber angle is fixed to $\phi = 4.15^{\circ}$ while each silica cone is 20 $\mu$m long, and \textbf{(d)} shows the supermode effective index along the propagation axis for different values of $L_{\mathrm{over}}$.}
\end{figure*}

We now present another way to use conical fibers in the absence of an on-chip bus waveguide: the fibers are placed in contact side to side, close to their tip, creating a fiber-to-fiber junction as depicted in Fig.~\ref{fig:coupling_f2f}. This new configuration creates a bus waveguide conceptually close to a conventional fiber taper \cite{DingBaker2010TaperFiber}, but with important differences such as an augmented mechanical rigidity, of importance in vibrating environments. We implemented this configuration in two distinct pulse tube cryostats where two fiber holders are placed on each side of the sample to approach the fiber close to a GaAs disk resonator\cite{Wasserman2022}. The fiber holders are moved in the three directions by low temperature, vacuum-compatible piezo positionning stages.
The integration of a fiber-to-fiber junction in this environment required no modification of the cryogenic nanopositionning setup.

An imaging optical setup comprising a camera is mounted below the vacuum can of the fridge with a microscope objective inside the fridge. To visualize the sample in the cold, a set of piezo-positioners (Attocube company) allows the sample to be moved in the confocal plane of the objective. Another positioner allows precise focus by moving the microscope objective along the confocal axis.
An image obtained in situ at 4~K is shown in Fig.~\ref{fig:image_camera_4K} showing a GaAs disk with its GaAs bus nano-waveguide (bottom) and a fiber-to-fiber junction formed by two contacted conical fiber tapers (top).
In order to make sure that conical fibers are coplanar with the disk, we use the fact that fibers are thin enough to be placed first below the disk (the pedestal height is 1.8 $\mu$m). 
When the two fibers are coplanar and close to the disk, they are laterally brought in contact with one another. Again, van der Waals interactions guarantee their resilient contact, which allows then to move one fiber without losing the coupling between the two. It is thus easy to tune the gap distance between the disk resonator and the fiber-to-fiber junction. Similarly to a conventional fiber taper, this fiber-to-fiber junction evanescently couples light to the disk whispering gallery modes.
This coupling method has been used in our group for resonant optical spectroscopy measurements on disk resonator and photoluminescence at low temperature (from 13 mK to 4 K). It allows highly efficiency and stable coupling even with the pulse tube turned on. 

\subsection{Simulation of the fiber-to-fiber junction}
The simulation geometry is parametrized by the fibers tip angle $\phi$ and their overlap length $L_{\mathrm{over}}$ as depicted in Fig.~\ref{fig:coupling_f2f}. Note that we can tune these two parameters experimentally: the angle $\phi$ by changing the lifting speed during the fiber fabrication (see Appendix~\ref{app:fabrication}) and the contact length between the two fibers by moving them with the piezo positionning stages. For a specific angle, a simulation has been performed with the EME method, for a polarization of the electric field parallel to the junction plane, and an optimum overlap length $L_{\mathrm{over}} = 16$ $\mu$m has been found to reach a transmission of optical power $\eta >99\%$ (see Fig.~\ref{fig:simu_f2f_eff}). We also provide the value of the supermode effective index along the axis in Figure~\ref{fig:simu_f2f_neff}, which shows the adiabatic coupling between the two conical fibers. 
Since we are interested in two different wavelength ranges in our experiments (1500-1600 nm and 800-950 nm), we optimized $\phi$ and $L_{\mathrm{over}}$ for both cases (see Table~\ref{tab:params_opti_2lambda}). Additional simulations are shown in Appendix~\ref{app:additional_simulations}.
\begin{table}
\hskip-1.5cm
    \begin{tabular}{c|ccc}
    $\lambda$ (nm) & $\phi$ ($^{\circ}$) & $L_{\mathrm{over}}$ ($\mu$m) & $\eta$ (\%)\\
    \hline
    1550 & 4.15 & 16 & 99.1\\
    \hline
    850 & 4.15 & 8 & 99.4
    \end{tabular}
    \caption{Optimal parameters found by 3D EME simulation for the fiber-to-fiber junction at both wavelengths of interest.}
\label{tab:params_opti_2lambda}
\end{table}

\subsection{Resonant spectroscopy at low temperature with the junction}
\begin{figure}
\subfloat[]{
\includegraphics[height=0.3\textwidth]{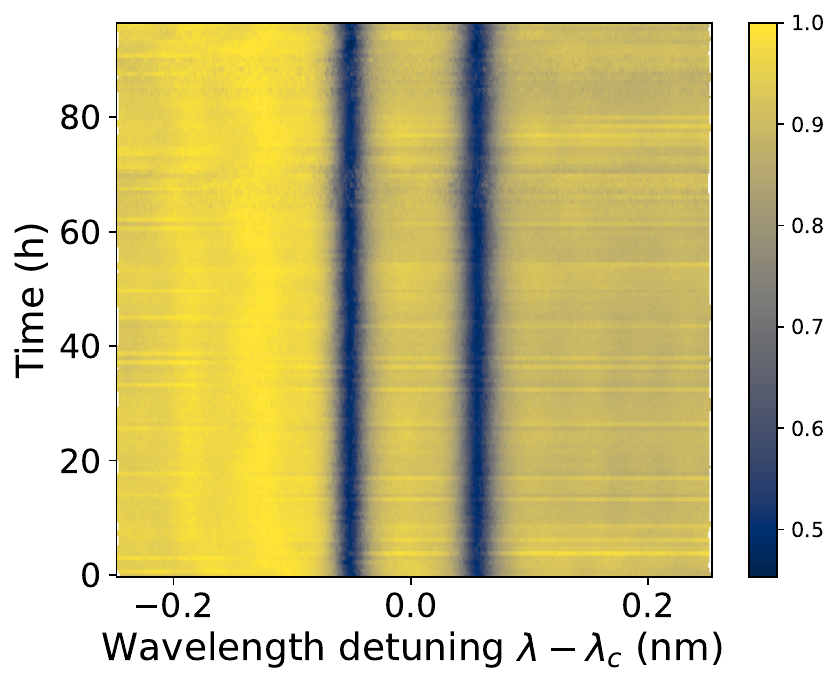}
\label{fig:spectro_telecom_mK}
}\\$^{\circ}$
\subfloat[]{
\includegraphics[height=0.3\textwidth]{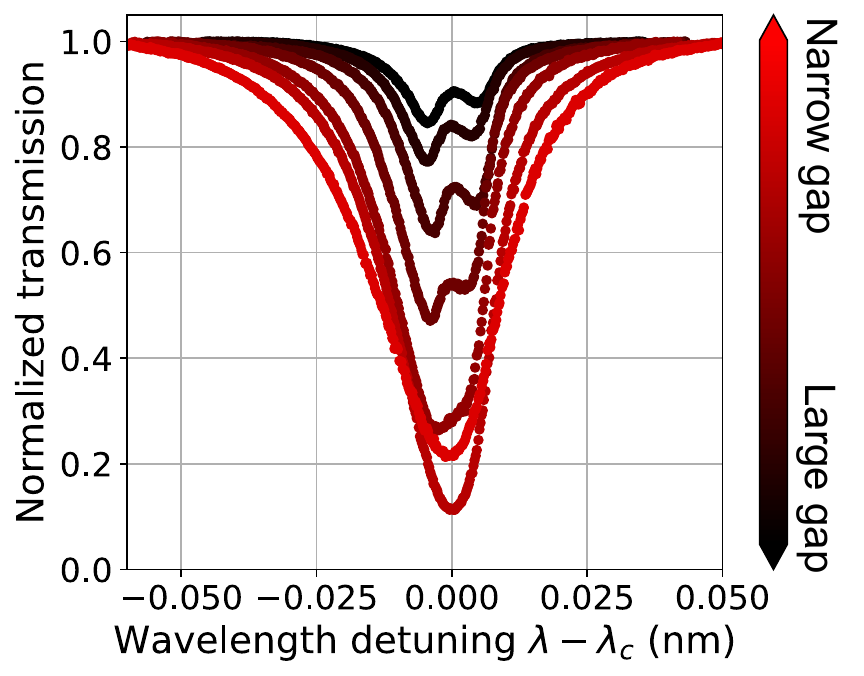}
\label{fig:spectro_telecom_RT}
}
\caption{
Optical transmission spectra in the telecom band, using a fiber-to-fiber junction evanescently coupled to the whispering gallery modes of a disk. \textbf{(a)} Normalized optical transmission spectra on a 1.3 $\mu$m radius, 320 nm thick disk resonator at millikelvin temperature. \textbf{(b)} Normalized optical transmission spectra on a 4.5 $\mu$m radius, 320 nm thick disk resonator at room temperature for different fiber junction-disk gaps. The spectra evolve from the presence of a doublet in the under-coupling regime (large gap, black) to a broader single Lorentzian line close to critical and over-coupling \cite{Ding2010LoopedFiber} (narrow gap, red).}
\label{fig:spectro_telecom}
\end{figure}
On top of its mechanical stability and easiness of alignment, the main experimental benefitsof the conical fiber junction is its high transmission of optical power, which is measured to be $\eta>95$\% at 1550 nm and $\eta >70$\% at 930 nm, for $\phi \approx 2^{\circ}$.
This is achieved without going through the painful task of optimizing an integrated GaAs waveguide geometry for each specific wavelength. Resonant spectroscopy measurements on a disk range were carried in the telecom (see Fig.~\ref{fig:spectro_telecom}). 
We were able to run experiments at mK temperatures during several days without experiencing any drift, as shown in Fig.~\ref{fig:spectro_telecom_mK}, which highlights the stability of this coupling method at low temperatures even in a dry cryostat setting with a pulse-tube running. 
The coupling between the fiber junction and the disk can be tuned by changing the position of the fibers: we observe a change in the contrast and linewidth of the optical resonances in Fig.~\ref{fig:spectro_telecom_RT}. 
We can attain both the over- and under coupling regime to the whispering gallery modes of the disk, as seen in the figure.

Similar measurements at 850 nm and 930 nm were carried at 4K on a 11 $\mu$m radius, 200 nm thick GaAs disk embedding InGaAs quantum wells, similar in dimensions to Fig~\ref{fig:fab}. 
Monomode fiber 780HP was employed for this purpose.
A key advantage of silica conical fibers in such measurements is to avoid the need of an integrated coupling waveguide that limits transmission because of the absorption close to transitions of the quantum well. With an integrated nano-waveguide embedding quantum wells and lensed fibers the maximum power transmission was $\eta = 0.1\%$.
Using a fiber-to-fiber junction instead, the measured transmission was $\eta>70\%$ in the 850-950 nm range.

\subsection{Photoluminescence measurement with conical fibers}

\begin{figure}
\subfloat[]{
\includegraphics[width=0.37\textwidth]{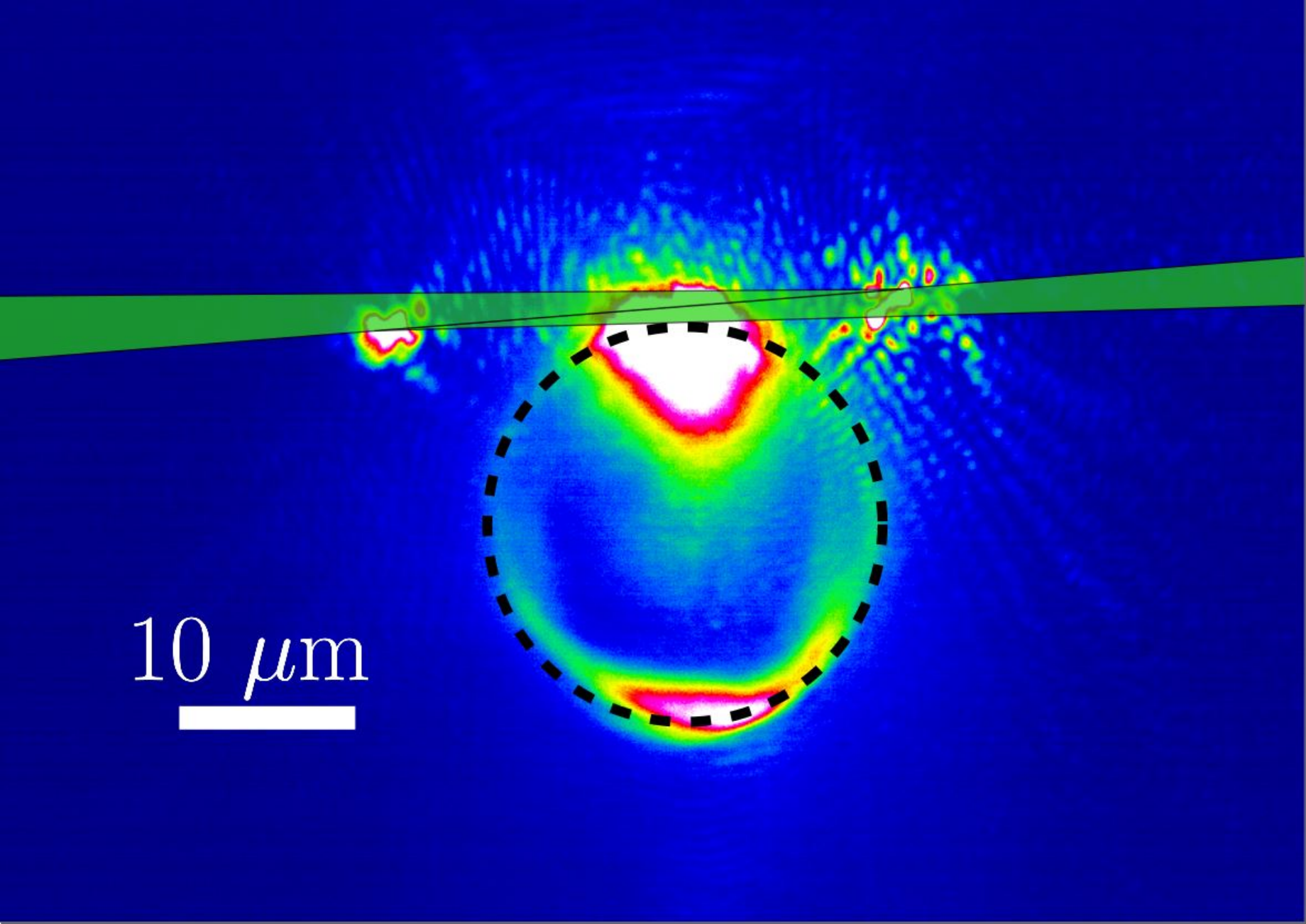}
\label{fig:PLmicroscope}
}\\
\subfloat[]{
\includegraphics[height=0.27\textwidth]{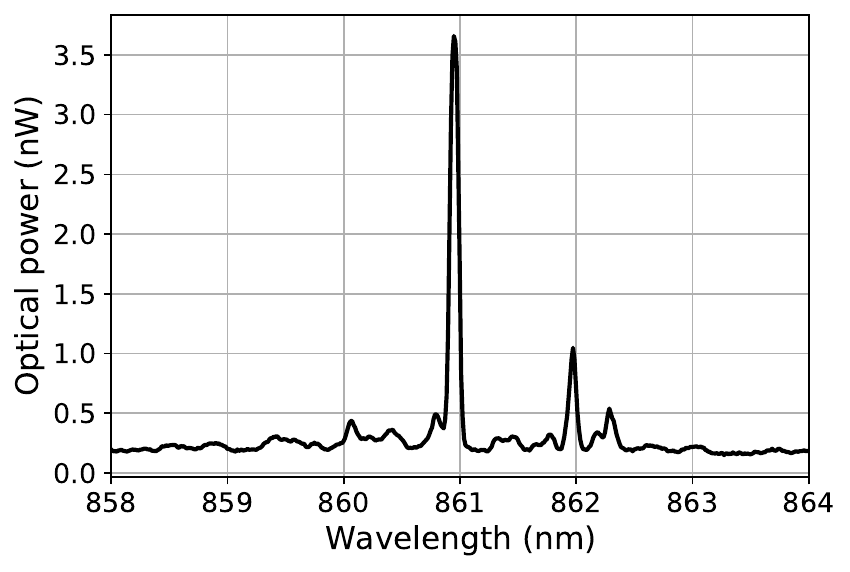}
\label{fig:PL}
}
\caption{\textbf{(a)} Photoluminescence of a disk resonator embedding InGaAs quantum wells imaged at 4K, when a non-resonant pump is employed. The dashed circle indicates the disk perimeter, and the conical fibers have been false colored in green. The disk radius and thickness are 11 $\mu$m and 200 nm respectively. \textbf{(b)} Corresponding photoluminescence spectrum, acquired on an optical spectral analyser connected at the output of the fiber.}
\end{figure}
Photoluminescence measurements can be carried with conical fiber tapers, by injecting a pump and collecting the light emitted by a GaAs disk embedding InGaAs quantum wells.
With an out-of-resonance pump (840 nm), the photoluminescence stems from the exciton reservoir and from exciton-polariton modes  \cite{Oliveira2023}, and can imaged directly by the silicon camera mounted in our confocal imagine system, as shown in Fig.~\ref{fig:PLmicroscope}. In Fig.~\ref{fig:PL}, we show that conical fibers can be used to collect and analyse this photoluminescence. Such measurement differs from confocal photoluminescence spectroscopy, where light is collected in the far field. The conical fiber here acts as a local antenna, collecting the emitted light in the near field. 

\section{Conclusion}
We investigated the technique of single-sided etched conical fibers for optical coupling to nanophotonic waveguide and resonators made-out of GaAs, a high refractive index material. We demonstrated its potential for experiments in liquid environments, comparing it to commercial lensed fibers and grating couplers approaches. We also introduced a novel coupling technique that uses two conical fibers joined together. We showed that this technique is well suited for low-temperature nanophotonics experiments in two wavelength ranges: telecom (1.5 $\mu$m) and near infrared (850-950 nm). The efficiency, reliability, stability and simplicity of implementation of this method makes it a powerful tool for demanding experiments.
Note finally that while already reaching high-efficiency in the present report, there are still margins of improvement for the method. The very low loss level required by quantum control or in-liquid applications will greatly benefit from these developments.

\begin{acknowledgments}
We acknowledge Andrea Gerini for help with simulations, Stephan Suffit, J\'er\'emie Schumann and Lorenzo Lazzari for help in setting up the conical fibers fabrication procedure, and Martial Nicolas and Batiste Janvier for the conception and fabrication of fiber holders for liquid applications. This work was supported by the European Research Council through the NOMLI project (770933).
\end{acknowledgments}

\appendix

\section{Fabrication of etched conical fibers}
\label{app:fabrication}

\begin{figure}
    \includegraphics[width=0.48\textwidth]{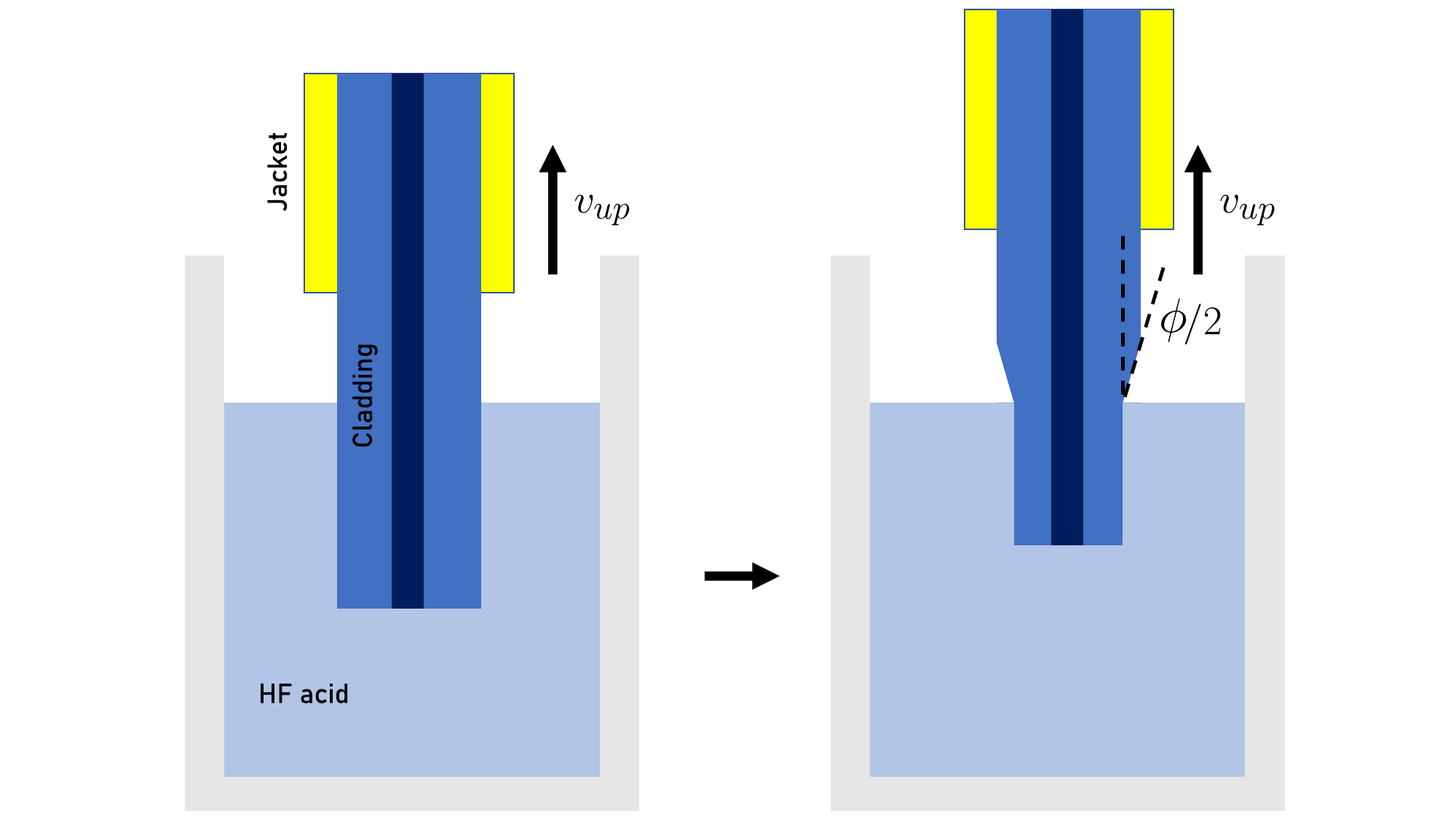}
    \caption{Fabrication principle of homemade etched conical fibers.}
\end{figure}
Single-mode fibers, typically Corning SMF-28 or 780HP are cleaved (or cut), stripped and carefully cleaned with acetone to dissolve the plastic jacket leftovers, then cleaned with isopropanol. The fibers are immersed in a 50\% HF solution before being slowly lifted out of the acid thanks to a vertical piezo positionner. By controlling the speed $v_{\mathrm{up}}$ at which the fibers are lifted, one can control the final cone angle $\phi$ according to the equation :
\begin{equation}
\tan{(\phi/2)} = \frac{R_{\mathrm{fiber}}}{v_{\mathrm{up}}T_{\mathrm{etch}}}
\end{equation}
where $R_{\mathrm{fiber}}$ is the non-etched, nominal fiber radius and $T_{\mathrm{etch}}$ is the time required to totally etch the fiber. For an SMF-28 fiber, $R_{\mathrm{fiber}}=62.5$ $\mu$m, and $T_{\mathrm{etch}}\approx$ 50 min at 20$^{\circ}$C. For a typical angle $\phi = 4^{\circ}$, the lift speed is $v_{\mathrm{up}}=500$ nm/s. After the chemical etching, we dip the conical fiber in isopropanol for 5 minutes.
With this method, the fabricated fibers exhibit a tip width on the order of 100 nm.

\section{Additional EME simulations}
\label{app:additional_simulations}

\begin{figure*}
\includegraphics[width=0.9\textwidth]{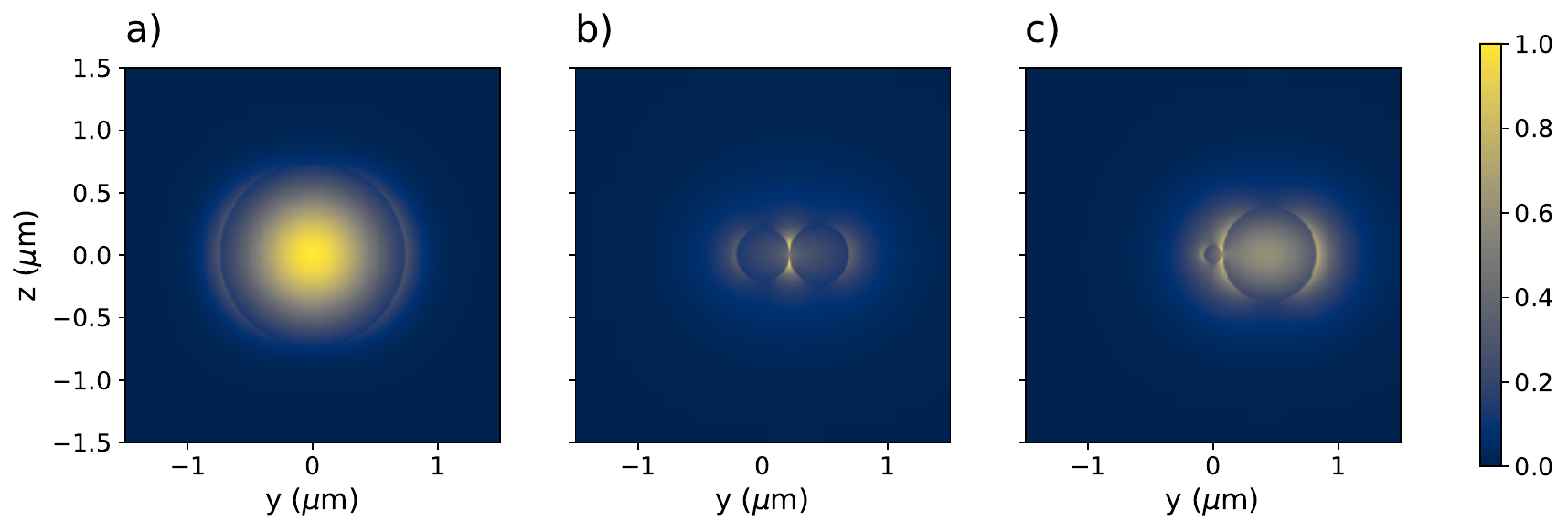}
\caption{Crossection of the normalized electric field modulus square $|E|^2$ for TE polarization at different locations along the fiber-fiber junction. The overlap length and fiber angle are $(L_{\mathrm{over}},\phi)=(15  \mu \text{m}, 4.15^{\circ})$.}
\label{fig:fielmaps_f2f}
\end{figure*}

\begin{figure*}
\subfloat[]{
\includegraphics[height=0.27\textwidth]{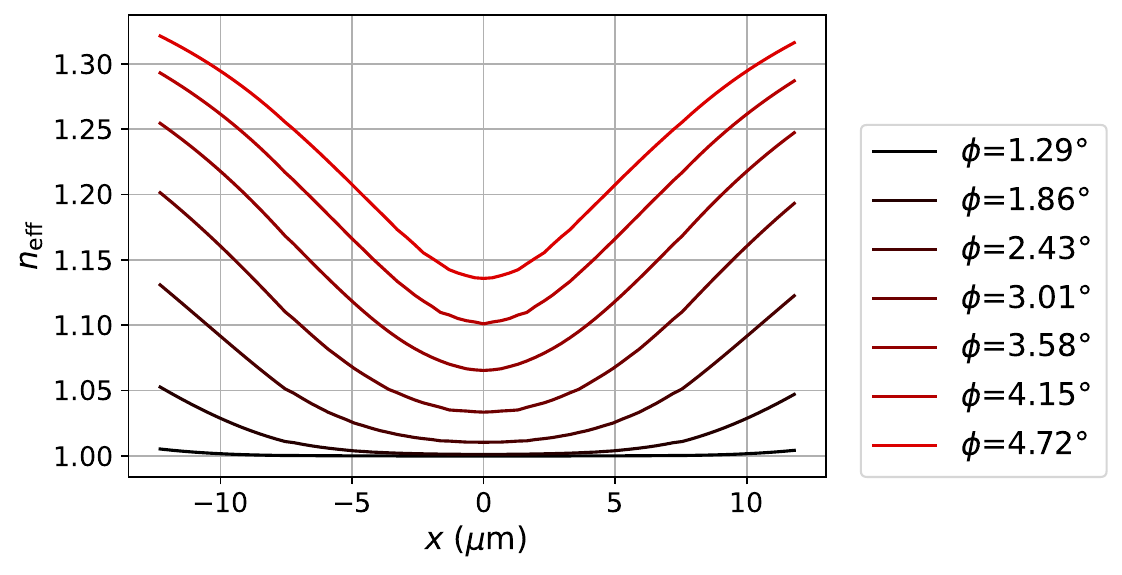}
\label{fig:simu_f2f_neff_vsR}
}
\subfloat[]{
\includegraphics[height=0.27\textwidth]{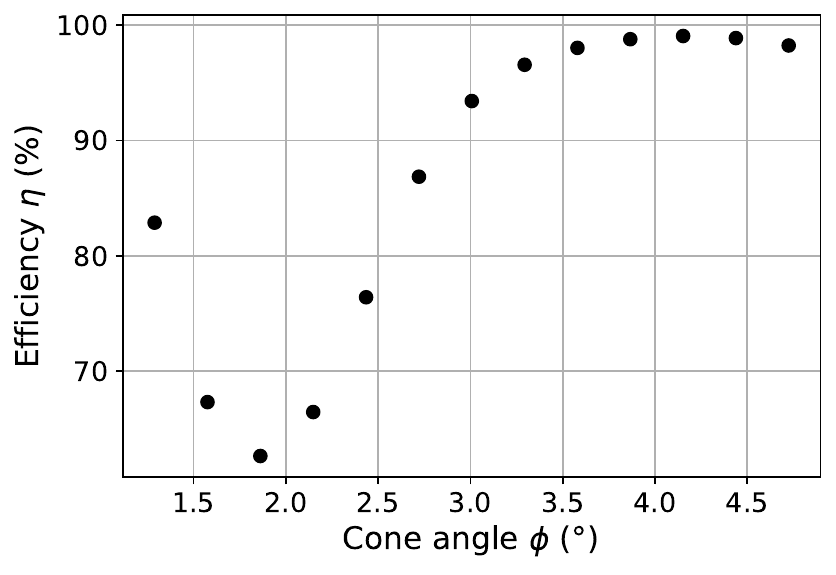}
\label{fig:simu_f2f_eff_vsR}
}
\caption{3D EME simulation of a conical fiber-to-fiber junction for different values of the fiber full angle $\phi$. \textbf{(a)} Fundamental supermode effective index $n_{\mathrm{eff}}$ along the light propagation axis for different values of $\phi$. \textbf{(b)} Power transmission efficiency $\eta$ as a function of $\phi$. For this simulation, $L_{\mathrm{over}} = 15$ $\mu$m and both fibers have the same cone angle.}
\label{fig:simus_f2f_vsR}
\end{figure*}

In this section we provide additional simulation results on the fiber-to-fiber junction configuration. Fig.~\ref{fig:fielmaps_f2f} shows the electric field supermode profile at different locations along the fiber-to-fiber junction. Note that the electric field density is high at the fibers interface in the middle of the junction, consistent with Fig.~\ref{fig:simu_f2f_neff} that shows that the effective index is minimum at this location. This could experimentally deteriorate $\eta$ because of roughness-induced scattering.
In Fig.~\ref{fig:simus_f2f_vsR}, we show simulations similar to Figs.~\ref{fig:simu_f2f_eff} and~\ref{fig:simu_f2f_neff}, but where $\phi$ is varied instead of $L_{\mathrm{over}}$. As expected, the larger the cone angle, the better the confinement of the supermode, even in the middle of the junction (see Fig.~\ref{fig:simu_f2f_neff_vsR}). The variation of $\eta$ as a function of $\phi$ (see Fig.~\ref{fig:simu_f2f_eff_vsR}) is in contrast less trivial: the evolution is non-monotonous and the angle $\phi \approx4^{\circ}$ provides an optimal efficiency.

\nocite{*}
\bibliography{aipsamp}

\end{document}